\documentclass[twocolumn,amsmath,amssymb,nofootinbib]{revtex4-2}

\usepackage{subfiles}
\usepackage{graphicx}
\usepackage{dcolumn}
\usepackage{bm}
\usepackage[
    colorlinks=true,
    citecolor=blue,
    linkcolor=blue,
    urlcolor=blue,
    pdfborder={0 0 0}
]{hyperref}
\usepackage{verbatim}
\usepackage{siunitx}
\usepackage{upgreek}
\usepackage{multirow}
\usepackage{subcaption}
\usepackage{caption}
\usepackage{ragged2e}
\graphicspath{{\subfix{Figures/}}}
\usepackage{comment}
\usepackage{rotating}
\usepackage{orcidlink}

\begin{document}

\title{SynradG4: A Geant4-Based Extension for Synchrotron Radiation Background Studies in the ePIC Detector at the Electron-Ion Collider}

\author{Andrii Natochii \orcidlink{0000-0002-1076-814X}}            %
 \email{natochii@bnl.gov}
\affiliation{Brookhaven National Laboratory, Upton, New York 11973, USA}

\date{\today}

\begin{abstract}
The Electron-Ion Collider (EIC) will operate at high luminosity with multi-GeV, high-current electron beams, resulting in substantial synchrotron radiation (SR) emission in the electron storage ring (ESR). A detailed understanding of SR photon transport in the complex three-dimensional interaction region (IR) geometry is critical for estimating backgrounds in the ePIC detector and for developing effective shielding and masking strategies.

This paper presents SynradG4, an EIC-specific extension of Geant4 for fast photon tracking in vacuum based on established models of SR reflection and rough-surface scattering. SynradG4 integrates the specular and diffuse photon reflection models of Synrad+ into the Geant4 geometry and field framework, while disabling all bulk matter interactions to achieve high-statistics transport through the 50-m-long IR vacuum system. The absorbed photons are then passed to a second-stage DD4hep simulation, in which electromagnetic processes such as photoabsorption, Compton scattering, Rayleigh scattering, and fluorescence are enabled for their propagation through the beam pipe and detector materials.

SynradG4 is not intended to replace general SR simulation codes; rather, it complements them by providing the workflow and geometry integration capabilities needed for EIC-specific background studies. Benchmark tests show that SynradG4 reproduces Synrad+ to within about 1\% for both specular and diffuse reflection, and that it agrees well with Synrad3D and with the native Geant4 X-ray reflection model for smooth surfaces. The differences observed for rough surfaces are traced to the convention used for the roughness attenuation factor (Geant4) and to the approximations of the parametrized diffuse-reflection model (Synrad3D). Using the full IR geometry and machine lattice, we present the first SR background estimates for the ePIC detector and evaluate the impact of potential SR masks.

\end{abstract}

\maketitle

\section{\label{sec:Introduction}Introduction}

The Electron-Ion Collider (EIC)~\cite{osti_1765663}, to be built at Brookhaven National Laboratory (BNL), will be a major step forward in our understanding of the fundamental structure of matter. Designed to probe quantum chromodynamics (QCD) in unprecedented detail, it will provide new insight into the behavior and dynamics of quarks and gluons, the elementary constituents of atomic nuclei. Its commissioning is planned for the 2030s.

The EIC will collide high-energy electron beams with protons and a wide range of ions, mapping the distributions of quarks and gluons within nucleons and nuclei with unprecedented precision. A central goal is to understand how gluons mediate the strong force, which confines quarks inside protons and neutrons and binds nucleons into nuclei.

The EIC will also address the origin of the nucleon spin, which remains unresolved despite decades of research. By colliding polarized beams, it will provide the data needed to determine the contributions of quarks and gluons to the proton spin.

In addition, the EIC will explore (i) dense gluon systems, (ii) matter under extreme conditions, and (iii) potentially new states of matter. Its flexibility in collision energy and ion species makes it a powerful tool for both targeted measurements and exploratory research, and the technological challenges of building the machine are expected to drive advances in accelerator and detector technologies.

Although the EIC is not a synchrotron light source, the synchrotron radiation (SR) emitted by the high-current, \SIrange{5}{18}{GeV} electron beam in the electron storage ring (ESR) is a significant source of beam-induced background. Hard X-rays generated upstream can be reflected at grazing angles along the vacuum system of the interaction region (IR) and reach the ePIC detector unless they are properly intercepted. Previous lepton colliders, such as LEP~\cite{Roudeau1994_SRmasks_LEP2}, HERA~\cite{Niebuhr2009_BackgroundAtHERA}, and SuperKEKB~\cite{LEWIS201969}, employed upstream SR masks and collimation systems, and similar strategies must be considered for the EIC. Given the complexity of the EIC IR geometry and the need for integration with the detector simulation based on the Detector Description Toolkit for High Energy Physics (DD4hep)~\cite{dd4hep}, dedicated modeling tools are essential.

The remainder of Section~\ref{sec:Introduction} gives an overview of the collider and the detector and discusses SR as a source of beam-induced background, its impact on the detector, and possible mitigation strategies. Section~\ref{sec:SectionA} reviews the codes available for SR simulation and the EIC-specific requirements that motivated the development of a new framework for synchrotron radiation simulation in Geant4 (SynradG4). Section~\ref{sec:SectionB} describes SynradG4, which uses the Geant4 toolkit~\cite{AGOSTINELLI2003250,1610988,ALLISON2016186} to simulate SR, with emphasis on the models of photon reflection at the vacuum--metal interface. Section~\ref{sec:SectionC} benchmarks SynradG4 against other SR simulation codes. Section~\ref{sec:SectionD} presents the resulting estimates of the SR background in the ePIC detector, which underline the importance of accurate SR simulations for background mitigation and detector optimization. Finally, Section~\ref{sec:Conclusion} summarizes our findings and outlines future plans.

\subsection{Collider design}

\begin{figure}[htbp]
\centering
\includegraphics[width=\linewidth]{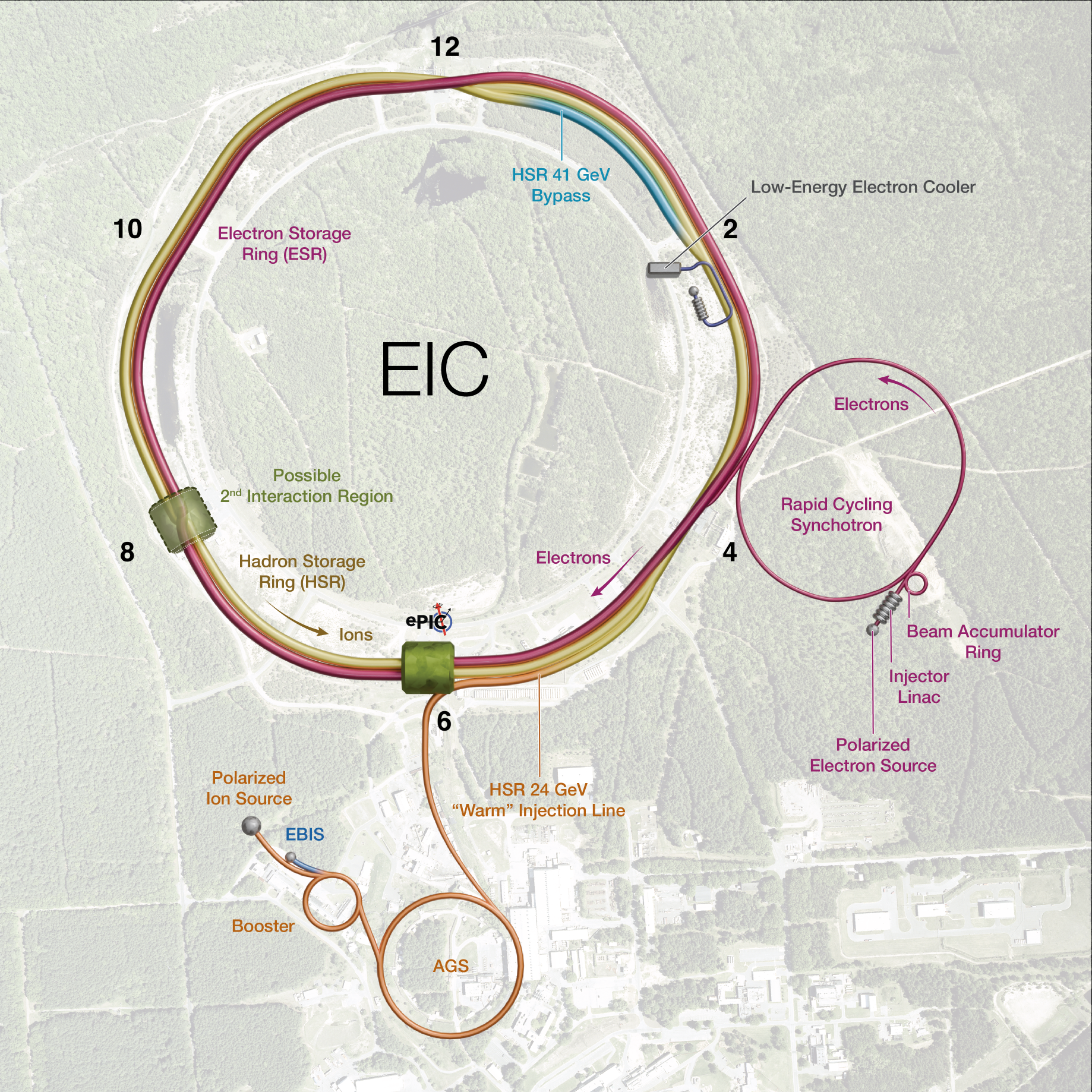}
\caption{\label{fig:EIC_schematic_drawing}Schematic layout of the EIC. The numbers 2 to 12 label the six straight sections around the rings.}

\end{figure}

The EIC will be built in the tunnel of the Relativistic Heavy Ion Collider (RHIC), which is about \SI{3.8}{km} in circumference. In the current design, one of the existing RHIC rings will be operated, with minimal modifications, as the Hadron Storage Ring (HSR), with beam energies from \SI{41}{GeV} to \SI{275}{GeV}. A new ESR will be installed in the same tunnel to store electron beams with energies of 5, 10, and \SI{18}{GeV}. Covering center-of-mass (CM) energies from 20 to \SI{140}{GeV}, the machine aims to reach an instantaneous luminosity of up to \SI{1e34}{cm^{-2}.s^{-1}} by colliding highly polarized electron and light-ion beams with a time-averaged polarization of about 70\%. Electrons will also be collided with a wide range of ions, from protons to uranium. The present project scope includes one IR and one detector, with a possible future upgrade to two IRs (Fig.~\ref{fig:EIC_schematic_drawing}). Table~\ref{tab:MachineParameters} lists the main machine parameters for electron--proton operation.

\begin{table*}[htbp]
\centering
    \caption{\label{tab:MachineParameters}
    EIC beam parameters adapted from Table~3.3 of Ref.~\cite{osti_1765663}. The quantities $\beta^{*}$ and $\Delta p/p$ denote the betatron function at the interaction point and the fractional momentum spread, respectively.}
    \begin{tabular}{l|cc|cc|cc}
    \hline\hline
    Species & proton & electron & proton & electron & proton & electron \\
    Energy [GeV] & 275 & 18 & 275 & 10 & 100 & 5 \\
    \hline
    Bunch intensity [$10^{10}$] & 19.1 & 6.2 & 6.9 & 17.2 & 4.8 & 17.2 \\
    No. of bunches & \multicolumn{2}{c|}{290} & \multicolumn{2}{c|}{1160} & \multicolumn{2}{c}{1160} \\
    Beam current [A] & 0.69 & 0.227 & 1 & 2.5 & 1 & 2.5 \\
    RMS emittance, H/V [nm] & 18/1.6 & 24/2 & 11.3/1 & 20/1.3 & 26/2.3 & 20/1.8 \\
    RMS bunch length [cm] & 6 & 0.9 & 6 & 0.7 & 7 & 0.7 \\
    RMS $\Delta p/p$ [$10^{-4}$] & 6.8 & 10.9 & 6.8 & 5.8 & 9.7 & 6.8 \\
    $\beta^{*}$, H/V [cm] & 80/7.1 & 59/5.7 & 80/7.2 & 45/5.6 & 61/5.5 & 78/7.1 \\
    Luminosity [$\mathrm{10^{33}~cm^{-2}s^{-1}}$] & \multicolumn{2}{c|}{1.54} & \multicolumn{2}{c|}{10} & \multicolumn{2}{c}{3.68} \\
    \hline\hline
    \end{tabular}
\end{table*}

\subsection{Detector design}

\begin{figure}[htbp]
\centering
\includegraphics[width=\linewidth]{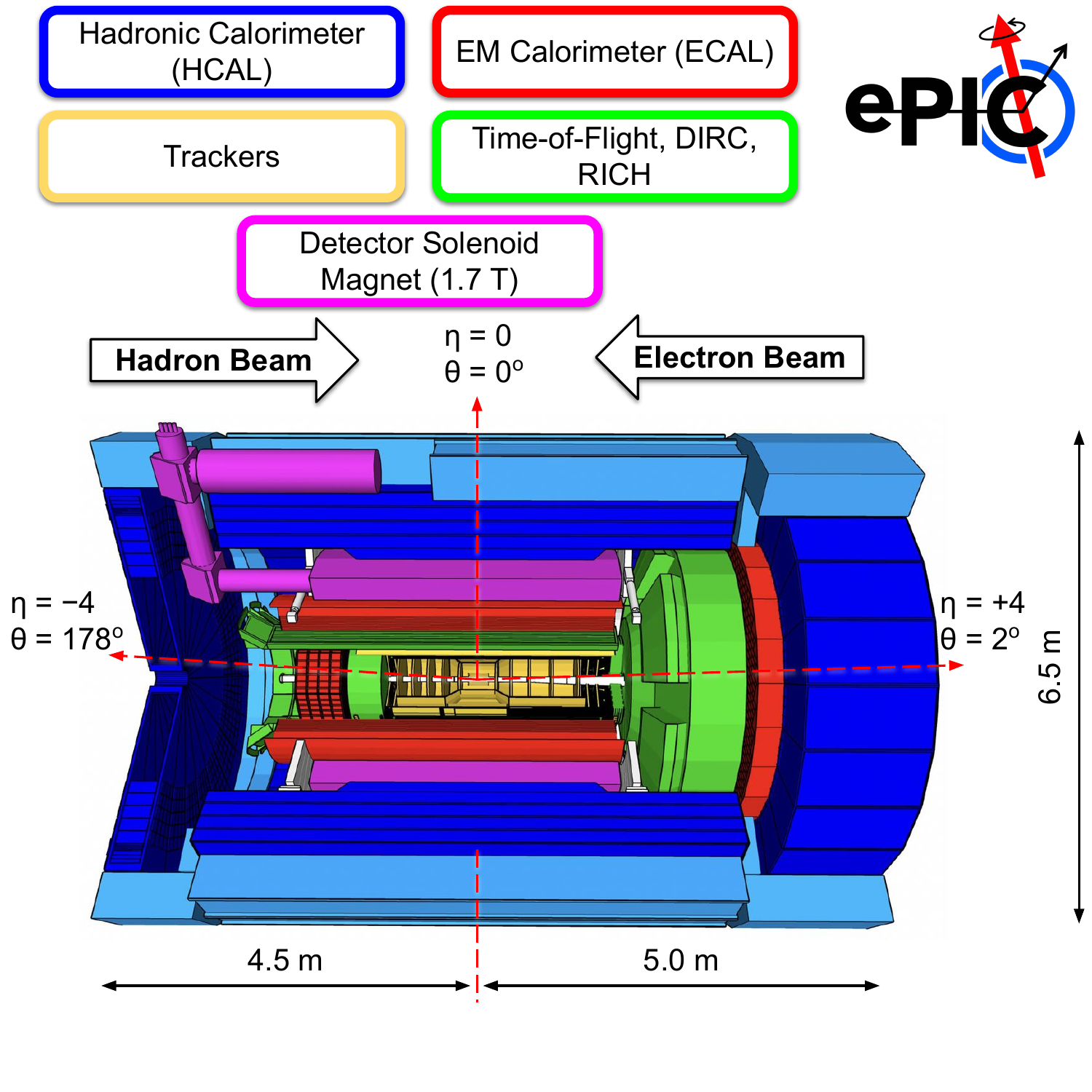}
\caption{\label{fig:ePIC_drawing}Schematic drawing of the ePIC detector; $\theta$ and $\eta$ denote the polar angle and the pseudorapidity, respectively.}

\end{figure}

The ePIC detector~\cite{ePIC_web-page} will be installed in IR6, the straight section at the 6~o'clock position (Fig.~\ref{fig:EIC_schematic_drawing}). It comprises vertexing and tracking, particle identification (PID), and calorimetry subsystems, as shown in Fig.~\ref{fig:ePIC_drawing}.

\begin{figure}[htbp]
\centering
\includegraphics[width=\linewidth]{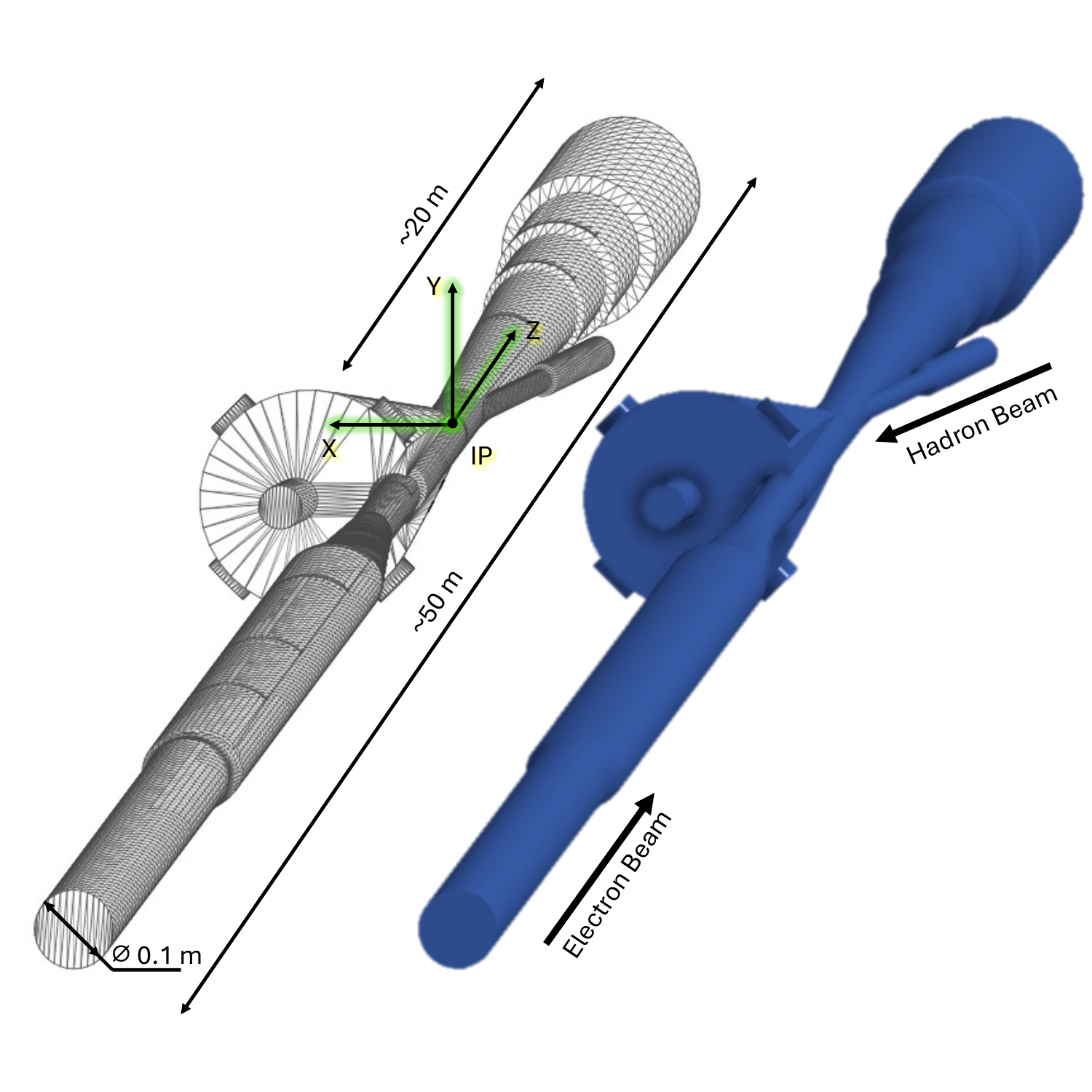}
\caption{\label{fig:IR_beampipe}Drawing of the IR vacuum beam pipe around the IP.}

\end{figure}

Figure~\ref{fig:IR_beampipe} shows the central vacuum beam pipe of the ESR. The section around the interaction point (IP), referred to as the IP beam pipe, has an inner diameter of \SI{62}{mm} and extends to \SI{80}{cm} and \SI{67}{cm} from the IP on the backward (electron-going) and forward (hadron-going) sides, respectively. It is made of about \SI{800}{\micro m}-thick beryllium with a \SI{5}{\micro m}-thick gold coating on the inner surface. Outside the IP region, up to the ends of the superconducting cryostats (about $\SI{20}{m}$ from the IP on both sides), the beam pipe is made of \SI{2}{mm}-thick stainless steel coated with a \SI{75}{\micro m} copper layer. The rest of the beam pipe is made of \SI{3}{mm}-thick copper.

The ePIC silicon vertex and tracking detector~\cite{LI2023168687}, based on \SI{65}{nm} monolithic active pixel sensor (MAPS) technology, provides a low material budget of about 0.05\%~$X/X_\mathrm{0}$ per layer and a high spatial resolution with a \SI{10}{\micro m} pitch. To improve the angular resolution of the tracks and thereby the separation of hadrons ($\pi$, $K$) in the PID detectors, micro-pattern gaseous detectors (MPGDs) based on $\upmu$RWELL and Micromegas technologies are used in the central tracker.

The time-of-flight (ToF) subsystem uses AC-coupled low-gain avalanche diode (AC-LGAD) technology~\cite{LI2023168687} to identify low-momentum charged particles ($3\sigma$ $\pi/K/p$ separation up to \SI{2.4}{GeV/c}) with high precision in the barrel and forward regions. In the central region, a high-performance detector of internally reflected Cherenkov light (hpDIRC)~\cite{kalicy2022developinghighperformancedircdetector} extends the momentum coverage, with $3\sigma$ $\pi/K/p$ separation up to \SI{6}{GeV/c}. In the hadron-going endcap, high-momentum coverage ($3\sigma$ $\pi/K/p$ separation up to \SI{50}{GeV/c}) is provided by the dual-radiator ring-imaging Cherenkov (dRICH) detector~\cite{VALLARINO2024168834}. In the backward region, the proximity-focusing ring-imaging Cherenkov (pfRICH) detector~\cite{pfRICH_CDR} provides charged-particle identification with $3\sigma$ $\pi/K/p$ separation up to \SI{7}{GeV/c}.

Electromagnetic calorimeters (ECALs)~\cite{ECAL_EIC} with high energy resolution ($< 10\% / \sqrt{E}$) surround the ePIC vertex and central subsystems to detect the scattered lepton, provide lepton identification against a large hadron background, and detect particles produced in semi-inclusive processes. Hadron calorimeters (HCALs) with an energy resolution of $< 50\% / \sqrt{E}$ are designed for jet energy measurements and for the reconstruction of deep-inelastic scattering (DIS) kinematics.

\subsection{Beam-Induced Background}

One of the main challenges in operating the EIC will be the control of beam-induced backgrounds in the detector. Among them, SR from the ESR is particularly important because of the high energy and current of the electron beam.

SR is emitted when relativistic charged particles, such as electrons, are deflected by magnetic fields. In the EIC, high-current (up to \SI{2.5}{A}) electron beams with energies of $\sim \SI{10}{GeV}$ will be guided around the ring by magnetic fields of $\sim \SI{0.1}{T}$. This inevitably produces intense SR of a few kW of power, which can reach the detector beam pipe and affect both the performance and the lifetime of the sensitive ePIC subsystems.

\subsubsection{Characteristics and Impact}

SR from the ESR has a continuous spectrum extending from infrared to X-ray wavelengths. Its intensity and spectrum depend on the beam energy and current, the bending radius, and the magnetic field strength. The primary concerns associated with SR at the EIC are:

\begin{itemize}
    \item \textbf{Detector Noise}: SR photons produce background hits in the detectors, which can obscure the signals of interest from electron--ion collisions.

    \item \textbf{Radiation Damage}: Prolonged exposure to SR can cause radiation damage to sensitive detector components, including the sensors and the readout electronics.

    \item \textbf{Heat Load}: The power deposited by SR can result in significant heat loads on detector and beam-pipe components.
\end{itemize}

\subsubsection{Mitigation Strategies}

Several strategies to mitigate the impact of SR are under investigation for the EIC:

\begin{itemize}
    \item \textbf{Shielding}: Carefully designed shielding, such as SR masks made of strongly absorbing materials and with a dedicated geometry, intercepts SR before it reaches sensitive detector regions. The masks must be placed strategically to minimize the radiation reaching critical components.

    \item \textbf{Collimators}: Collimators help to control the transverse distribution of the beam and therefore the spread of the emitted SR. Constraining the beam to well-defined trajectories reduces the radiation that reaches the detectors.

    \item \textbf{Advanced Detector Materials}: Radiation-hard materials and components extend the operational lifetime of the detectors.

    \item \textbf{Cooling Systems}: Efficient cooling is essential to remove the heat load generated by SR, especially in the superconducting final-focus magnets close to the IP.
\end{itemize}

Experience from previous high-energy lepton colliders (e.g., LEP, HERA, and SuperKEKB) shows that upstream SR masks and collimation systems are essential for attenuating the hard X-ray flux into the interaction region. The EIC IR adds complexity through its asymmetric geometry, tight magnet apertures, and coated beam-pipe surfaces.

Because masks and absorbers intercept the direct SR fan, the photons that reach the detector region are predominantly those that have been reflected, once or several times, by the beam-pipe walls. Their angular distribution after reflection is therefore as important as the reflection probability itself. For a rough surface, part of the reflected intensity is scattered diffusely, away from the specular direction~\cite{beckmann1963scattering,PhysRevSTAB.18.040704}, and diffusely reflected photons can enter regions that are geometrically shadowed from both the direct and the specularly reflected radiation. A dedicated study for the present EIC IR geometry and masking configuration~\cite{NatochiiPotekhin2026} showed that, compared with a specular-only treatment, the inclusion of diffuse reflection substantially increases the photon population reaching the IP beam pipe, including photons with energies above \SI{1}{keV}, owing to both increased photon survival and angular redistribution. A consistent treatment of specular and diffuse reflection is therefore required for reliable SR background predictions.

Together, these factors motivate a simulation framework that models specular and diffuse photon reflection and absorption on tens of thousands of facets and propagates the resulting photon field into a full detector simulation.

\section{\label{sec:SectionA}Synchrotron Radiation Simulation}

Several well-established tools exist for SR emission and scattering studies, most notably Synrad3D, Synrad+, and the X-ray reflection process of Geant4. Their direct use for the full EIC interaction region is, however, limited by workflow rather than by physics. For the EIC, the simulation must (i) import a $\sim\SI{50}{m}$ long CAD-based vacuum model containing $\sim 30000$~facets, (ii) record the absorbed-photon coordinates simultaneously on all facets, (iii) use the actual ESR lattice files, and (iv) pass the absorbed-photon distributions to DD4hep for detector-level background estimates. SynradG4 was developed to meet these practical requirements while reusing the validated SR reflection and scattering physics of Synrad+.

Accurate SR simulations are essential for detector protection and performance optimization: they must provide detailed predictions of the radiation environment in the IR to guide the design and implementation of mitigation measures. SR simulations for the EIC, however, face several challenges:

\begin{itemize}
    \item \textbf{High Statistics and CPU Resources}: The large number of SR photons produced in the ESR magnets requires substantial computing power and access to high-performance computing facilities.

    \item \textbf{Accurate X-ray Reflection Modeling}: Predicting the radiation environment requires accurate models of the interaction of SR with the beam-pipe surfaces, including the material-dependent reflectivity and absorption and the angular distribution of specularly and diffusely reflected photons.

    \item \textbf{Complex Geometry and Boundary Conditions}: The intricate EIC IR, with curved beam trajectories and varying beam-pipe cross sections (Fig.~\ref{fig:IR_beampipe}), requires flexible geometry handling and careful setup.
\end{itemize}

\subsection{Background Study Procedure}

\begin{figure}[htbp]
\centering
\includegraphics[width=\linewidth]{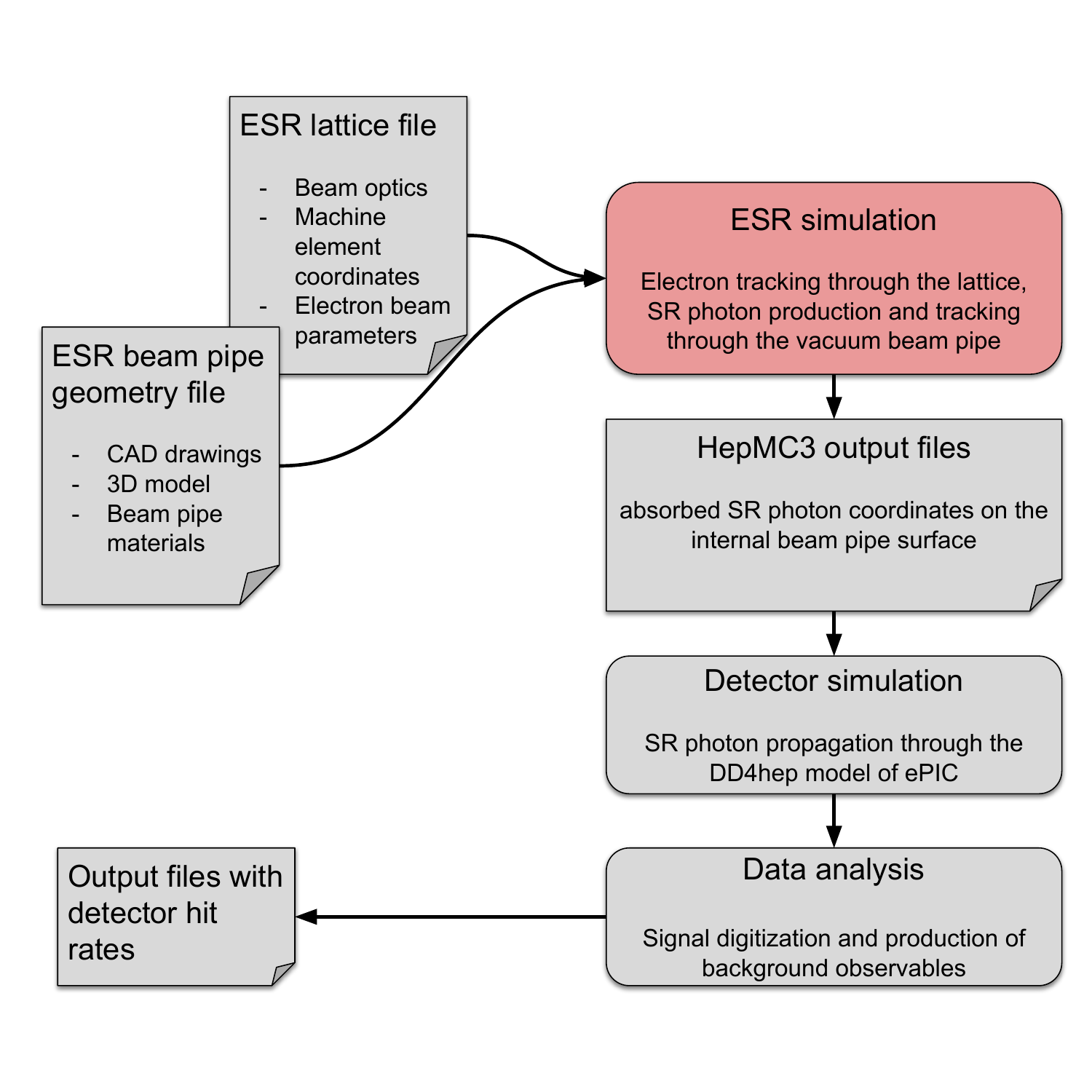}
\caption{\label{fig:SR_BG_study}Workflow of the SR background study.}

\end{figure}

The SR background study for the ePIC detector, shown schematically in Fig.~\ref{fig:SR_BG_study}, consists of the following steps:

\begin{enumerate}
    \item \textbf{ESR simulation}: The ESR lattice and beam-pipe geometry files are used to build the vacuum volume; high-energy electrons are tracked through the magnetic fields, and the SR photons they emit are transported until they are absorbed on the inner surface of the beam pipe.

    \item \textbf{Detector simulation}: The absorbed photons from the ESR simulation are propagated through the ePIC detector model with DD4hep, which is part of \textit{eic-shell}, the container environment developed for EIC/ePIC simulation and data analysis~\cite{eic_shell}. The detector response is recorded as hits with energy deposits above the calibration thresholds.

    \item \textbf{Data analysis}: The hit distributions in the detector are analyzed to obtain the hit rates corresponding to the experimental background observables.
\end{enumerate}

The most critical ingredient of the study, the ESR simulation (highlighted in red in Fig.~\ref{fig:SR_BG_study}), is discussed in detail in the following sections.

The simulation is thus performed in two stages. In Stage~1 (SynradG4), photons are transported only in vacuum, using Synrad+-based reflection and absorption models at the vacuum boundary, with all Geant4 bulk-matter interactions disabled. In Stage~2 (DD4hep), the absorbed photons are used as sources, and the full electromagnetic physics is enabled for their propagation through the beam pipe and the detector. This separation provides substantial performance gains while preserving the physical accuracy of the simulation.

\subsection{Available Simulation Tools}

Although SR simulations are of great interest for most $\rm e^{+}e^{-}$ machines, only a few frameworks can propagate electrons through magnetic fields, generate SR, and track the photons in vacuum. In this paper, we consider three of the most widely used codes for three-dimensional (3D) SR tracking, whose X-ray reflection models are supported by dedicated measurements.

\textit{Synrad3D}~\cite{etde_22538279}, developed at Cornell University, is built on the Bmad~\cite{SAGAN2006356} libraries and models the production of SR by electrons and its scattering from the vacuum-chamber walls, using an analytical model for diffuse scattering from surfaces with finite roughness~\cite{PhysRevAccelBeams.20.020708}. The shape of the vacuum chamber is represented by a sequence of sub-chambers, each with its own cross section in the plane transverse to the beam axis. The probability of specular (mirror-like) reflection from a rough surface is computed with the expression of Ref.~\cite{beckmann1963scattering}, while the diffuse reflection model is based on Kirchhoff (scalar) diffraction theory~\cite{beckmann1963scattering,JAOgilvy_1987}. Its predictions were compared with X-ray scattering measurements on technical vacuum-chamber surfaces in Ref.~\cite{PhysRevSTAB.18.040704}.

\textit{Synrad+}~\cite{synradPlus}, developed at CERN, transports SR photons through complex vacuum geometries imported as polygon meshes. It offers a wide range of options for the surface properties of the beam-pipe walls; the probability of specular reflection from a rough surface is described by the Debye--Waller factor~\cite{PhysRevSTAB.18.040704}. The Synrad+ diffuse scattering model follows the Synrad3D approach~\cite{PhysRevAccelBeams.20.020708}, but instead of evaluating the full expression, which involves an infinite sum, it uses the high-energy limit of the scattering distribution and approximates the resulting angular distributions by Gaussian functions, which greatly speeds up the computation~\cite{Ady:2157666}. The Synrad+ results presented in this paper were obtained with release version~1.5.10.

\textit{Geant4}~\cite{AGOSTINELLI2003250,1610988,ALLISON2016186} has provided, since release~11.2.0~\cite{geant4rel}, the G4XrayReflection process for the reflection of X-rays at the boundary between a low-density and a high-density material~\cite{geant4xray}. The process computes the specular reflectivity of a smooth surface from the optical constants of Ref.~\cite{HENKE1993181} and accounts for surface roughness through a Névot--Croce attenuation factor~\cite{NevotCroce}. In this paper, \textit{Geant4-11.2.0} refers to this X-ray reflection model.

Synrad3D and Geant4-11.2.0 use the same X-ray data for smooth surfaces, from the LBNL database~\cite{HENKE1993181}, whereas Synrad+ uses an extended library of reflectivity tables for common materials~\cite{Ady:2157666}.

While Synrad3D and Synrad+ contain accurate, experimentally validated models for SR reflection from rough surfaces, integrating them into a full detector-simulation chain is challenging in the EIC case. Synrad3D describes vacuum chambers by cross-sectional slices and does not natively import large tessellated CAD geometries. Synrad+ can import STL meshes, but its design, in which absorbed photons are scored per facet, requires sequential facet-by-facet runs for high-statistics scoring on $\sim 30000$~facets. Geant4 describes the specular component of X-ray reflection, including its attenuation by surface roughness, but does not yet include diffuse scattering from rough surfaces~\cite{NatochiiPotekhin2026}. SynradG4 therefore integrates the established Synrad+ reflection physics into Geant4, enabling a single-pass, EIC-specific workflow that handles the full beam-pipe geometry and interfaces directly with DD4hep.

\section{\label{sec:SectionB}New Framework: SynradG4}

SynradG4 is a Geant4-based extension designed for fast and accurate SR photon transport in the vacuum of the EIC IR. To maximize computational efficiency, SynradG4 imports only the vacuum volume of the IR beam pipe from its STL tessellated geometry and does not construct any material volumes. All bulk electromagnetic interactions relevant for SR photons with energies up to hundreds of keV (photoelectric absorption, Compton and Rayleigh scattering, and fluorescence) are therefore deferred to the Stage~2 detector simulation in DD4hep, where the full Geant4 electromagnetic physics lists are applied. Within SynradG4, only boundary processes are enabled: the Synrad+-based rough-surface reflection model, with its specular and diffuse components, and photon absorption at the vacuum--metal interface. This approach preserves the validated surface behavior while enabling very fast X-ray tracking over about \SI{50}{m} of vacuum, with multiple reflections on $\sim 30000$~facets.

When a photon is absorbed on the inner surface, SynradG4 records its position, energy, and direction. These data are used in the Stage~2 DD4hep simulation, in which the full beam-pipe materials and all electromagnetic processes are included. Because SR photons at grazing incidence interact primarily with the topmost surface layer, a single-layer reflectivity model is sufficient for Stage~1, while fluorescence and scattering in the bulk material are treated in Stage~2.

\subsection{\label{sec:ReflectionModel}Reflection model}

\begin{figure}[htbp]
\centering
\includegraphics[width=\linewidth]{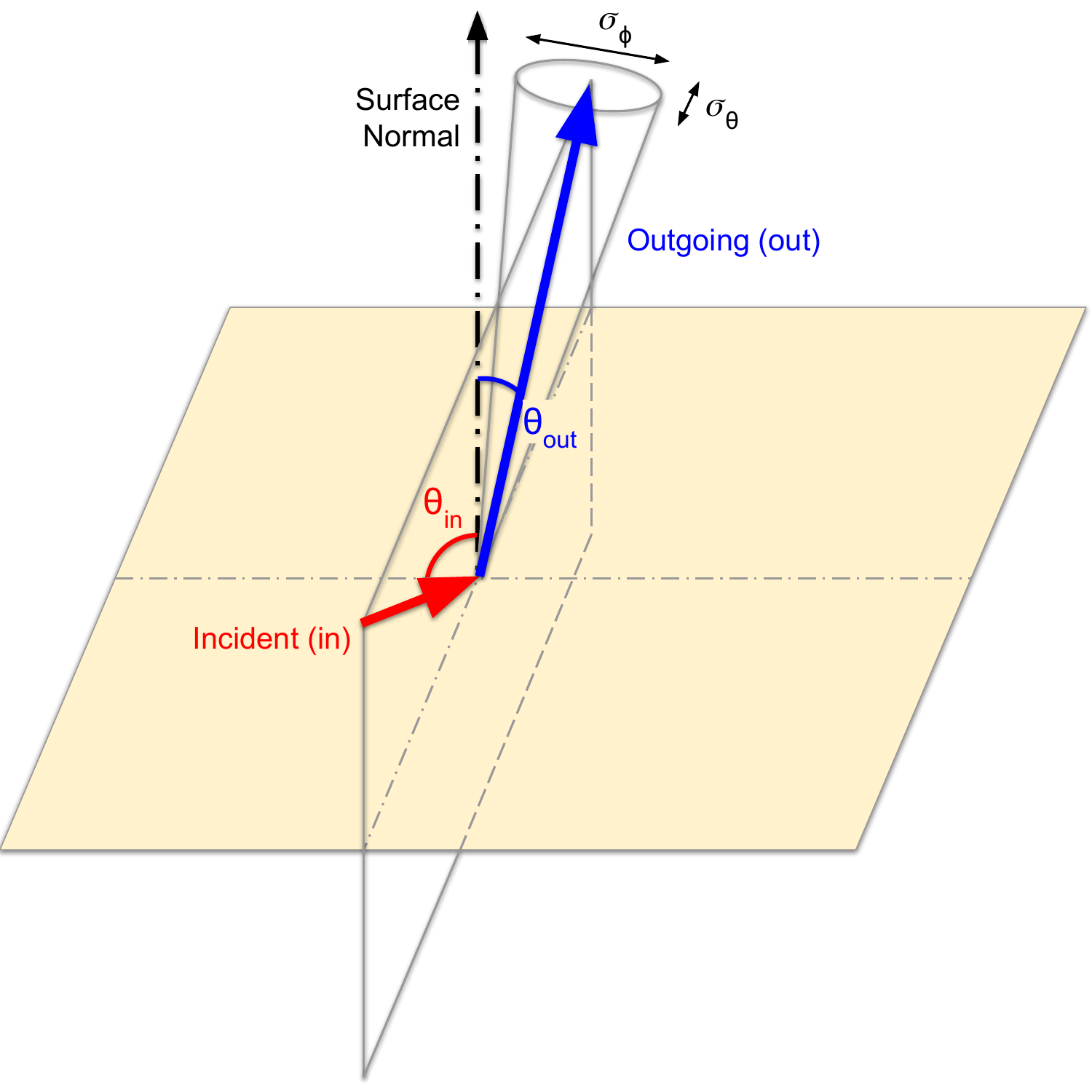}
\caption{\label{fig:ReflectionAngles}Notation for the diffuse reflection angles in SynradG4. The polar angles $\theta_\mathrm{in}$ and $\theta_\mathrm{out}$ are measured from the surface normal; $\sigma_\mathrm{\theta}$ and $\sigma_\mathrm{\phi}$ are the widths of the distributions of the polar and azimuthal (out-of-plane) angles around the specular direction.}

\end{figure}

SR photon reflection at the beam-pipe--vacuum interface is modeled by a dedicated discrete process, \textit{GammaReflectionProcess}\footnote{In earlier versions of this preprint, the exponent of Eq.~(\ref{eq:eq0}) was a factor of two smaller, the diffuse reflection angles were obtained by perturbing the polar and azimuthal angles of the reflected direction in the global coordinate system, and the coefficients of Eqs.~(\ref{eq:eq1}) and (\ref{eq:eq2}) were taken from Ref.~\cite{Ady:2157666}. The benchmark results in Section~\ref{sec:SectionC} were obtained with the revised implementation described here~\cite{SynradG4code}.}. The reflectivity of a smooth surface, $R_\mathrm{F}(E_\mathrm{\gamma},\theta_\mathrm{in})$, which depends on the photon energy $E_\mathrm{\gamma}$ and on the incidence angle $\theta_\mathrm{in}$ measured from the surface normal (Fig.~\ref{fig:ReflectionAngles}), is taken from the extended Synrad+ library, with logarithmic interpolation in energy and angle between the tabulated points. The surface roughness is characterized by its RMS height $\sigma$ (denoted $\sigma_\mathrm{RMS}$ in Section~\ref{sec:SectionC}) and its autocorrelation length $T$~\cite{beckmann1963scattering}. The probability that a reflected photon leaves the surface in the specular direction is given by the Debye--Waller factor of the specularly reflected intensity~\cite{beckmann1963scattering,PhysRevSTAB.18.040704,Ady:2157666}:
\begin{equation}
    P_\mathrm{spec} = \exp\!\left[-\left(2k_\mathrm{\perp}\sigma\right)^{2}\right] = \exp\!\left[-\left(\frac{4\pi\sigma\cos\theta_\mathrm{in}}{\lambda}\right)^{2}\right],
    \label{eq:eq0}
\end{equation}
where $k_\mathrm{\perp} = k\cos\theta_\mathrm{in}$ is the component of the photon wave vector normal to the surface, $k = 2\pi/\lambda = E_\mathrm{\gamma}/\hbar c$, $\lambda$ is the photon wavelength, and $\hbar c = \SI{197.327}{MeV.fm}$. For a smooth surface ($\sigma = 0$), $P_\mathrm{spec} = 1$, and photons are reflected specularly with probability $R_\mathrm{F}$.

Two treatments of rough surfaces are available in SynradG4:
\begin{itemize}
    \item In the \textit{specular} model, only specularly reflected photons are tracked, and the reflection probability is $R_\mathrm{F}\,P_\mathrm{spec}$. The remaining fraction of the reflected intensity, $R_\mathrm{F}(1-P_\mathrm{spec})$, which is scattered out of the specular direction, is treated as absorbed.
    \item In the \textit{diffuse} model, as in Synrad+~\cite{Ady:2157666}, the total reflection probability is $R_\mathrm{F}$: roughness redistributes the reflected photons between the specular and diffuse components without changing the total reflection probability. A reflected photon is reflected specularly with probability $P_\mathrm{spec}$ and diffusely otherwise.
\end{itemize}

For a diffuse reflection, the outgoing direction is sampled with the parametrization of Ref.~\cite{Ady:2157666}. In the local frame of the surface (Fig.~\ref{fig:ReflectionAngles}), the polar angle $\theta_\mathrm{out}$, measured from the surface normal, and the azimuthal angle $\phi_\mathrm{out}$ about the normal, with $\phi_\mathrm{out} = 0$ in the plane of incidence, are drawn from Gaussian distributions centered on the specular direction ($\theta_\mathrm{out} = \theta_\mathrm{in}$, $\phi_\mathrm{out} = 0$) with standard deviations
\begin{equation}
    \sigma_\mathrm{\theta} = \frac{2.9264}{\tau},
    \label{eq:eq1}
\end{equation}
\begin{equation}
    \sigma_\mathrm{\phi} = \frac{1}{\tau}\left(2.80657\, \theta_\mathrm{in}^{-1.00238}-1.00293\,\theta_\mathrm{in}^{1.22425}\right),
    \label{eq:eq2}
\end{equation}
where $\tau = T / \sigma$ and the angles are expressed in radians. As in Synrad+, the distribution of $\theta_\mathrm{out}$ is truncated to the interval $[0, \pi/2]$, so that the reflected photon always leaves the surface into the vacuum. The coefficients in Eqs.~(\ref{eq:eq1}) and (\ref{eq:eq2}) are those of the current Synrad+ reflection code~\cite{AdySynradReflection}; they differ from the values 2.9267 and 1.2266 quoted in Table~2.4 of Ref.~\cite{Ady:2157666} by less than 0.2\%, which changes $\sigma_\mathrm{\phi}$ by at most a few percent at grazing incidence. The sampled angles are converted into a direction in the global coordinate system using the local orthonormal frame formed by the surface normal and the projection of the incident direction onto the surface, so that the result does not depend on the orientation of the geometry.

The Gaussian parametrization approximates the angular distribution derived from Kirchhoff theory in Refs.~\cite{PhysRevSTAB.18.040704,PhysRevAccelBeams.20.020708} in its high-energy limit, $g = (4\pi\sigma\cos\theta_\mathrm{in}/\lambda)^2 \gg 1$. Apart from small deviations of the out-of-plane distribution for nearly grazing incidence ($1.55 < \theta_\mathrm{in} < \pi/2$)~\cite{Ady:2157666}, it reproduces this distribution well, at a small fraction of the computational cost of the full expression.

\subsection{\label{sec:G4Roughness}Comparison with the Geant4 roughness factor}

In the X-ray reflection model of Geant4-11.2.0, the reflectivity of a smooth surface, computed from the Henke optical constants~\cite{HENKE1993181}, is multiplied by the Névot--Croce attenuation factor~\cite{geant4xray,NevotCroce}
\begin{equation}
    f_\mathrm{NC} = \exp\!\left(-2k_{iz}k_{jz}\sigma^{2}\right),
    \label{eq:NC}
\end{equation}
where $k_{iz} = k_\mathrm{\perp}$, $k_{jz} = (1-\delta)\,k_{iz}$~\cite{geant4xray}, and $\delta$ is the decrement of the real part of the refractive index, $n = 1-\delta+i\beta$.

Roughness factors of this type can be written either for the amplitude of the reflected wave or for the reflected intensity. Following Névot and Croce~\cite{NevotCroce}, the Fresnel reflection coefficient $r_\mathrm{F}$ of a rough interface is multiplied by $\exp(-2k_{iz}k_{jz}\sigma^{2})$, where $k_{jz}$ is the normal component of the (complex) wave vector in the material, so that
\begin{equation}
    r = r_\mathrm{F}\, e^{-2k_{iz}k_{jz}\sigma^{2}} \quad \text{and} \quad R = |r|^{2} = R_\mathrm{F}\, e^{-4k_{iz}\,\mathrm{Re}(k_{jz})\,\sigma^{2}}.
    \label{eq:amplitude}
\end{equation}
For $k_{jz} = k_{iz}$, the intensity factor coincides with the Debye--Waller factor of Eq.~(\ref{eq:eq0})~\cite{Sinha1988}. The exponent of Eq.~(\ref{eq:NC}) has the amplitude form of Eq.~(\ref{eq:amplitude}), whereas Synrad+~\cite{Ady:2157666}, the analysis of Ref.~\cite{PhysRevSTAB.18.040704}, and SynradG4 use the intensity form of Eq.~(\ref{eq:eq0}); a comparative study of roughness models for X-ray and extreme-ultraviolet reflectance can be found in Ref.~\cite{Esashi_21}. In the energy range in which roughness attenuation becomes significant for the benchmark in Section~\ref{sec:SectionC}, $\delta \ll 1$ (for copper, $\delta \approx 10^{-3}$ at \SI{1}{keV}, decreasing with energy~\cite{HENKE1993181}). Hence $k_{jz} \simeq k_{iz}$ in Eq.~(\ref{eq:NC}), whose exponent is then one half of that of Eq.~(\ref{eq:eq0}). For the same value of $\sigma$, Geant4-11.2.0 therefore attenuates the specular reflection less than SynradG4 and Synrad+; equivalently, the two descriptions coincide if the roughness used in Eq.~(\ref{eq:NC}) is $\sqrt{2}$ times larger than that used in Eq.~(\ref{eq:eq0}). This difference in convention has to be taken into account when roughness values are transferred between the codes, and it is the main origin of the differences between SynradG4 and Geant4-11.2.0 for rough surfaces in Section~\ref{sec:SectionC}.

\section{\label{sec:SectionC}Benchmark}

The purpose of the benchmark is not to assess the physics accuracy of the individual codes -- Synrad3D, Synrad+, and Geant4 are based on established physical models -- but to demonstrate that SynradG4 reproduces their reflection behavior for identical geometry and surface parameters while enabling a single-pass, high-statistics workflow, and to understand the origin of any remaining differences.

\begin{figure*}[htbp]
\centering
\subfloat[\label{fig:Synrad3Dgeometry}Synrad3D geometry. Top: $X$--$S$ beam coordinate system. Bottom: $X_\mathrm{glob.}$--$Z_\mathrm{glob.}$ global coordinate system.]{\includegraphics[width=0.33\linewidth]{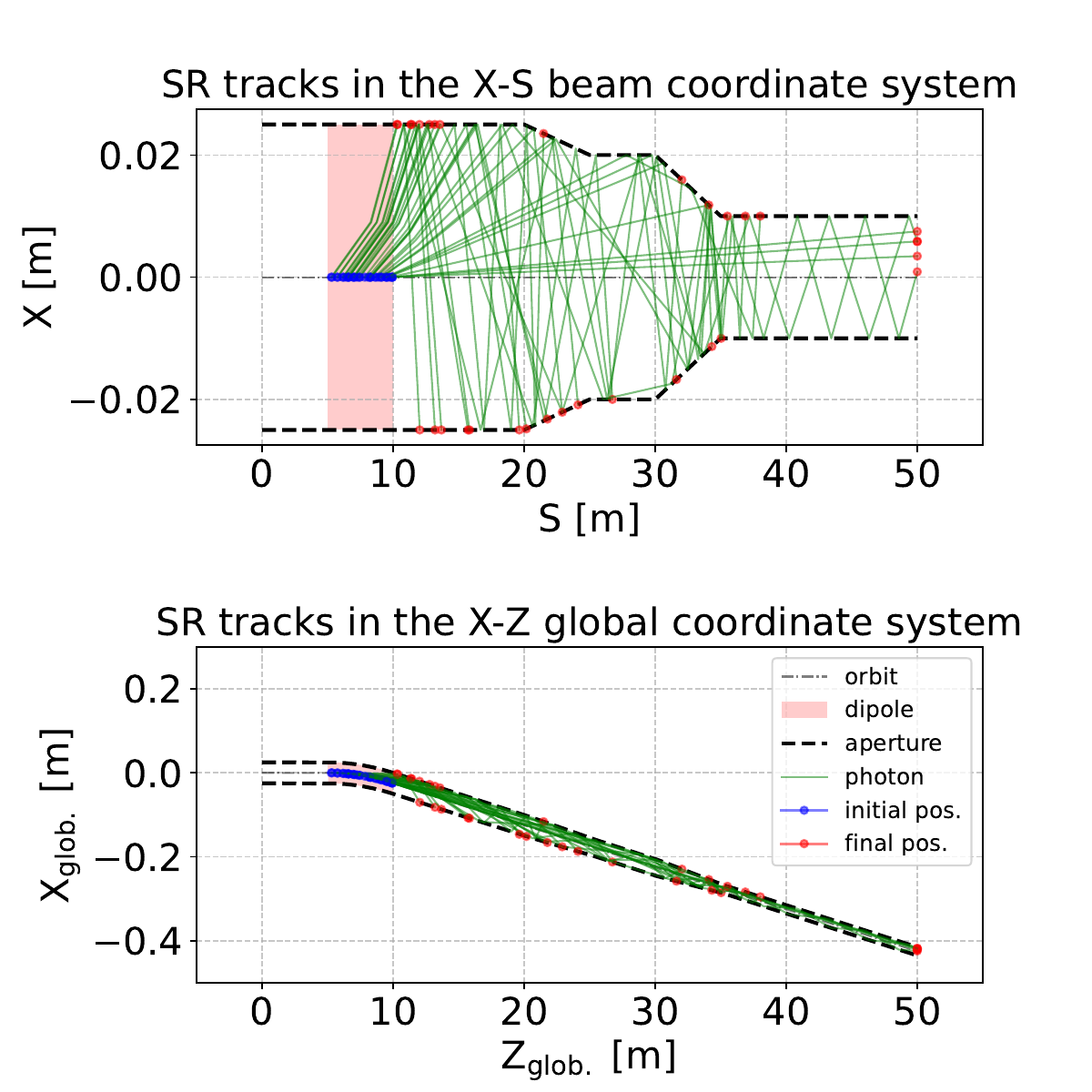}}
\hfill
\subfloat[\label{fig:SynradPgeometry}Synrad+ geometry.]{\includegraphics[width=0.33\linewidth]{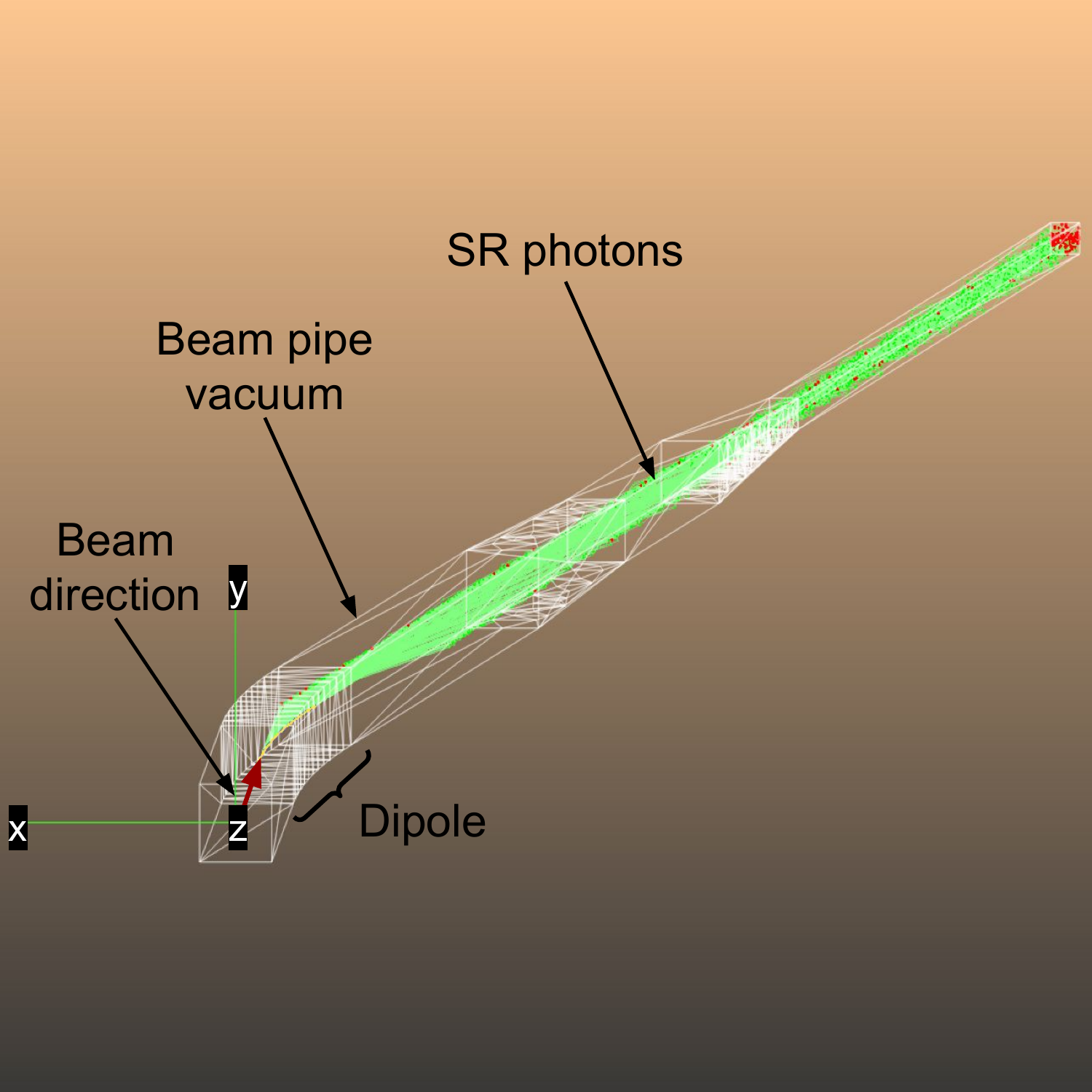}}
\hfill
\subfloat[\label{fig:SynradG4geometry}Geant4 geometry used for both SynradG4 and Geant4-11.2.0.]{\includegraphics[width=0.33\linewidth]{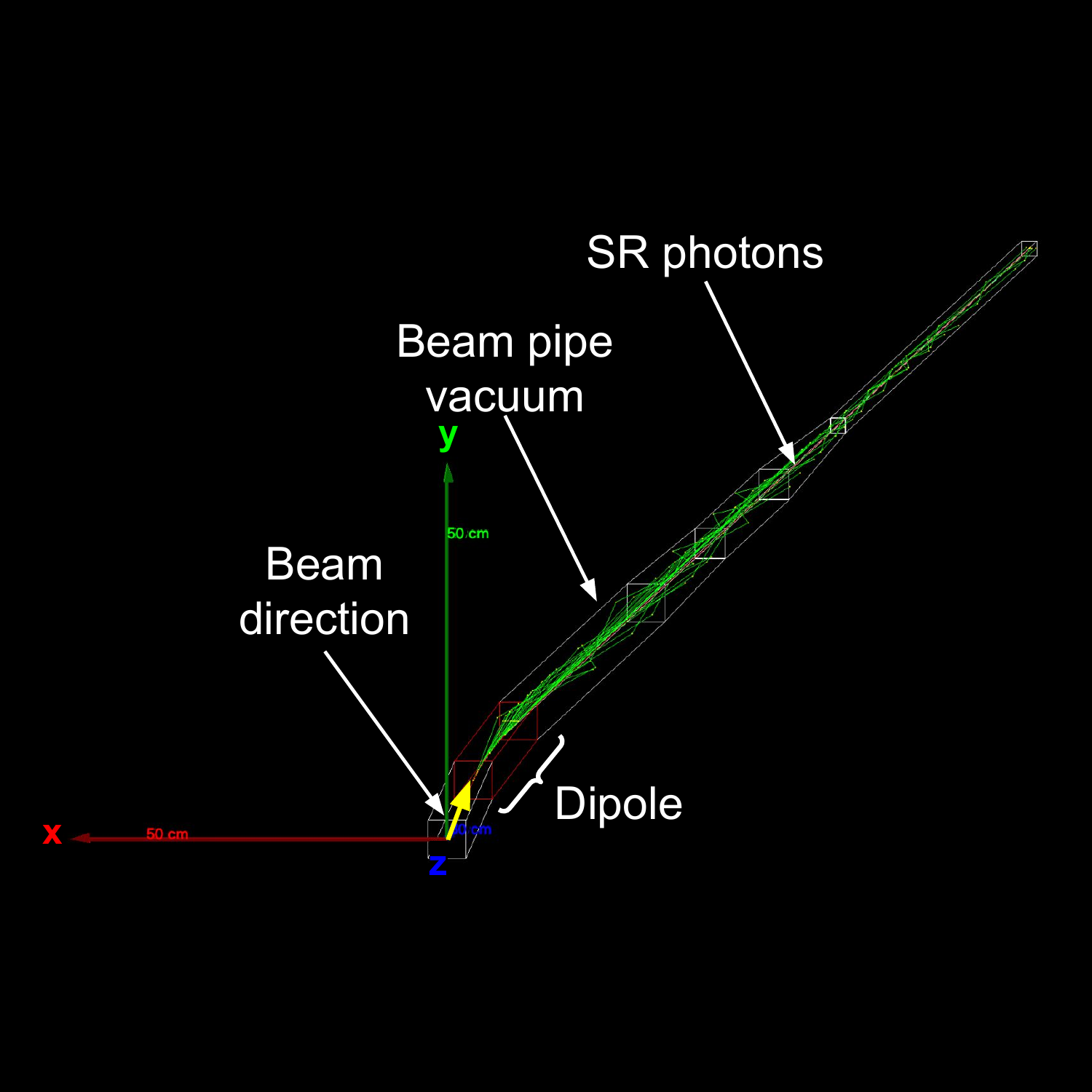}}
\caption{\label{fig:Geometry}Simulation models used for the benchmark.}

\end{figure*}

To benchmark the code, we compared SynradG4 with Synrad+, Synrad3D, and Geant4-11.2.0 for different surface roughness parameters and photon reflection models in a simple geometry: a \SI{50}{m} long vacuum beam pipe that includes a \SI{5}{m} long dipole magnetic field with a bending angle of \SI{10}{mrad} for \SI{18}{GeV} electrons.

\subsection{Setup}

Figure~\ref{fig:Synrad3Dgeometry} shows the Synrad3D model of the vacuum volume, represented by a series of cross-sectional planes with varying apertures in the plane transverse to the beam axis. The beam-pipe aperture in the vertical and horizontal planes is shown by black dashed lines, and the region of uniform dipole field is highlighted in red. SR photons are generated along the beam path within the dipole field, as indicated by the blue markers. After multiple reflections on the beam-pipe walls, the photon tracks, shown in green, end at the absorption points marked in red.

Because the ($X$, $S$) coordinates are curved inside the bending magnet, the photon trajectory between two hit points is not a straight line in the $X$--$S$ coordinate system of Synrad3D, as shown in Fig.~\ref{fig:Synrad3Dgeometry} (top). We therefore transformed all trajectories into the global $X_\mathrm{glob.}$--$Z_\mathrm{glob.}$ coordinate system, as illustrated in Fig.~\ref{fig:Synrad3Dgeometry} (bottom).

Figure~\ref{fig:SynradPgeometry} shows the 3D model of the same vacuum beam pipe in Synrad+, represented as a set of facets (white). The SR photon tracks produced in the dipole field region are shown in green, and the reflection and absorption points on the walls are marked in red.

Finally, Fig.~\ref{fig:SynradG4geometry} shows the 3D model of the vacuum beam pipe simulated in Geant4 and used by both SynradG4 and Geant4-11.2.0. The SR photon tracks generated in the dipole field region (red volume) are shown in green; after multiple reflections, the photons are absorbed on the beam-pipe walls (white volume).

\begin{table}
    \centering
    \begin{tabular}{l|c|c|c}
    \hline\hline
         Framework &  Setup-1 &  Setup-2 & Setup-3 \\
    \hline
         \multirow{2}{*}{Synrad3D}  &  \multirow{2}{*}{SRO = TRUE} & \multirow{2}{*}{\textit{N/A}} & $\sigma_\mathrm{RMS} = 50$,\\
                                    & & & $T = 10$\\
    \hline
         \multirow{2}{*}{Synrad+}   &  \multirow{2}{*}{RSS = OFF} &  \multirow{2}{*}{\textit{N/A}} & $\sigma_\mathrm{RMS} = 50$,\\
                                    &  & &$T = 10$\\
    \hline
         \multirow{2}{*}{SynradG4}  & $\sigma_\mathrm{RMS} = 0$ &  $\sigma_\mathrm{RMS} = 50$ & $\sigma_\mathrm{RMS} = 50$,\\
                                    & (spec.) &  (spec.) & $T = 10$ (diff.)\\
    \hline
         G4-11.2.0 & $\sigma_\mathrm{RMS} = 0$ &  $\sigma_\mathrm{RMS} = 50$ & \textit{N/A} \\
    \hline\hline
    \end{tabular}
    \caption{\label{tab:BenchmarkSetup}Simulation setups for the SynradG4 benchmark. The surface roughness $\sigma_\mathrm{RMS}$ and the autocorrelation length $T$ are given in nanometers and micrometers, respectively; ``spec.'' and ``diff.'' denote the specular and diffuse models of SynradG4 (Section~\ref{sec:ReflectionModel}).}
\end{table}

To compare the absorbed-photon distributions after multiple reflections along the vacuum beam pipe, we selected the straight section between $Z_\mathrm{glob.} = \SI{40}{m}$ and $Z_\mathrm{glob.} = \SI{45}{m}$. Depending on the functionality of each code, we studied the setups listed in Table~\ref{tab:BenchmarkSetup}. In Setup-1 and Setup-2, the codes simulate specular reflection with a surface roughness $\sigma_\mathrm{RMS}$ of \SI{0}{nm} and \SI{50}{nm} (benchmark value), respectively. In Setup-2, roughness reduces the specular reflection probability according to Eq.~(\ref{eq:eq0}) in SynradG4 and Eq.~(\ref{eq:NC}) in Geant4-11.2.0. Setup-3 models diffuse reflection with $\sigma_\mathrm{RMS} = \SI{50}{nm}$ and an autocorrelation length $T = \SI{10}{\micro m}$ ($\tau = 200$). Since SynradG4 can switch between its specular (spec.) and diffuse (diff.) reflection models, it covers all three setups.

\textit{N/A} denotes options that are not available: Synrad3D and Synrad+ do not provide a specular-only treatment with roughness attenuation, and Geant4-11.2.0 does not include diffuse reflection. In Synrad3D, the parameter SRO (specular reflection only) is set to TRUE, so that photons are always reflected specularly (the default is FALSE)~\cite{synrad3d_manual}. In Synrad+, the option RSS (rough surface scattering) is set to OFF, so that surface roughness is ignored and photons are reflected specularly without attenuation (the default is ON).

\begin{figure}[htbp]
\centering
\includegraphics[width=\linewidth]{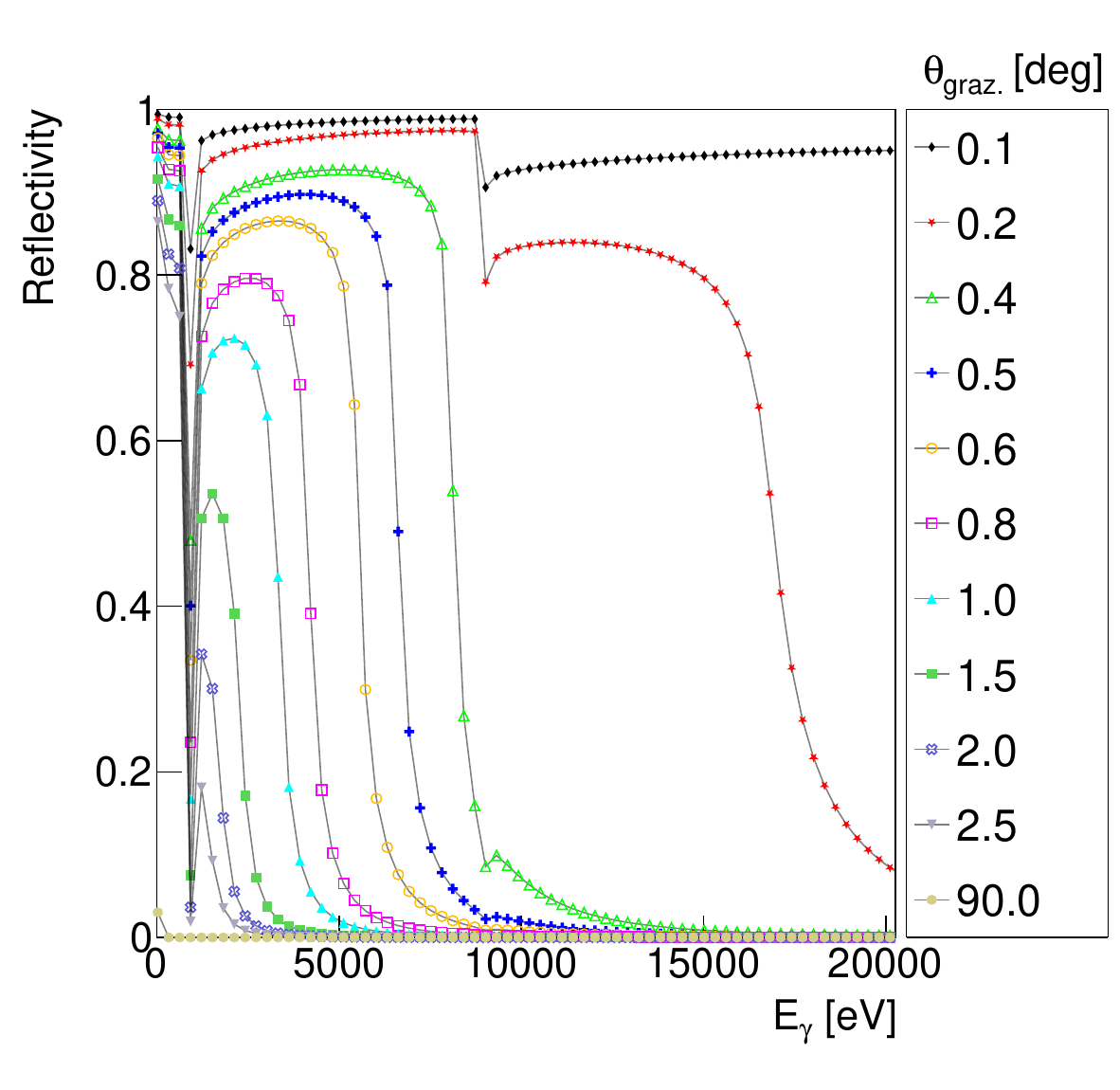}
\caption{\label{fig:Reflectivity}Specular reflectivity of a smooth copper surface as a function of the photon energy $E_\mathrm{\gamma}$ and the grazing angle $\theta_\mathrm{graz.} = \pi/2 - \theta_\mathrm{in}$, measured from the surface. Data from Ref.~\cite{HENKE1993181}.}

\end{figure}

In all setups, the walls of the vacuum beam pipe are made of copper, whose smooth-surface reflectivity as a function of photon energy and grazing angle is shown in Fig.~\ref{fig:Reflectivity}. Since the LBNL data are available only above \SI{30}{eV}, all codes track SR photons with energies above this threshold.

\subsection{Results}

\begin{figure*}[htbp]
\centering
\subfloat[\label{fig:setup1}Setup-1.]{\includegraphics[width=0.33\linewidth]{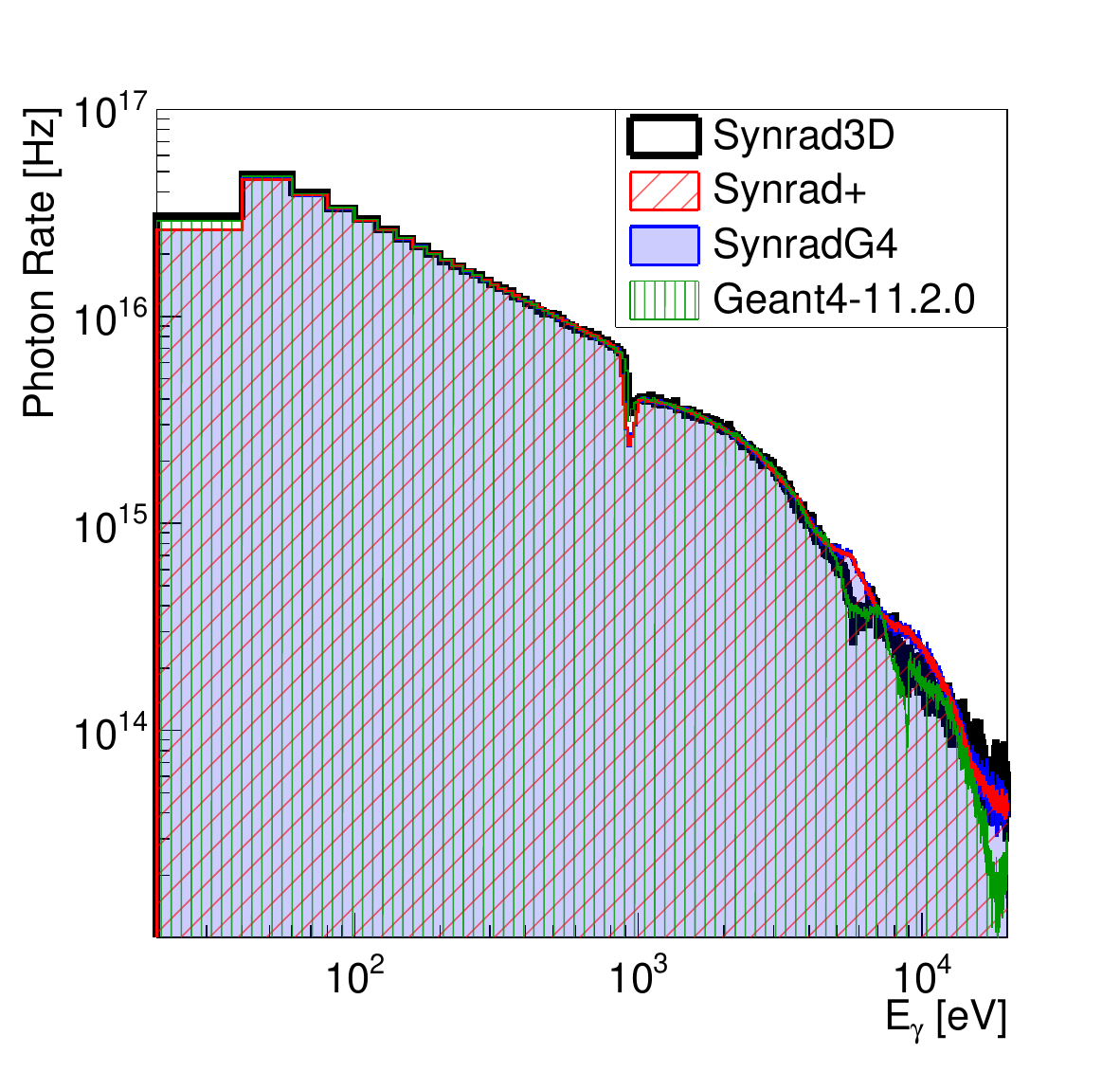}}
\hfill
\subfloat[\label{fig:setup2}Setup-2.]{\includegraphics[width=0.33\linewidth]{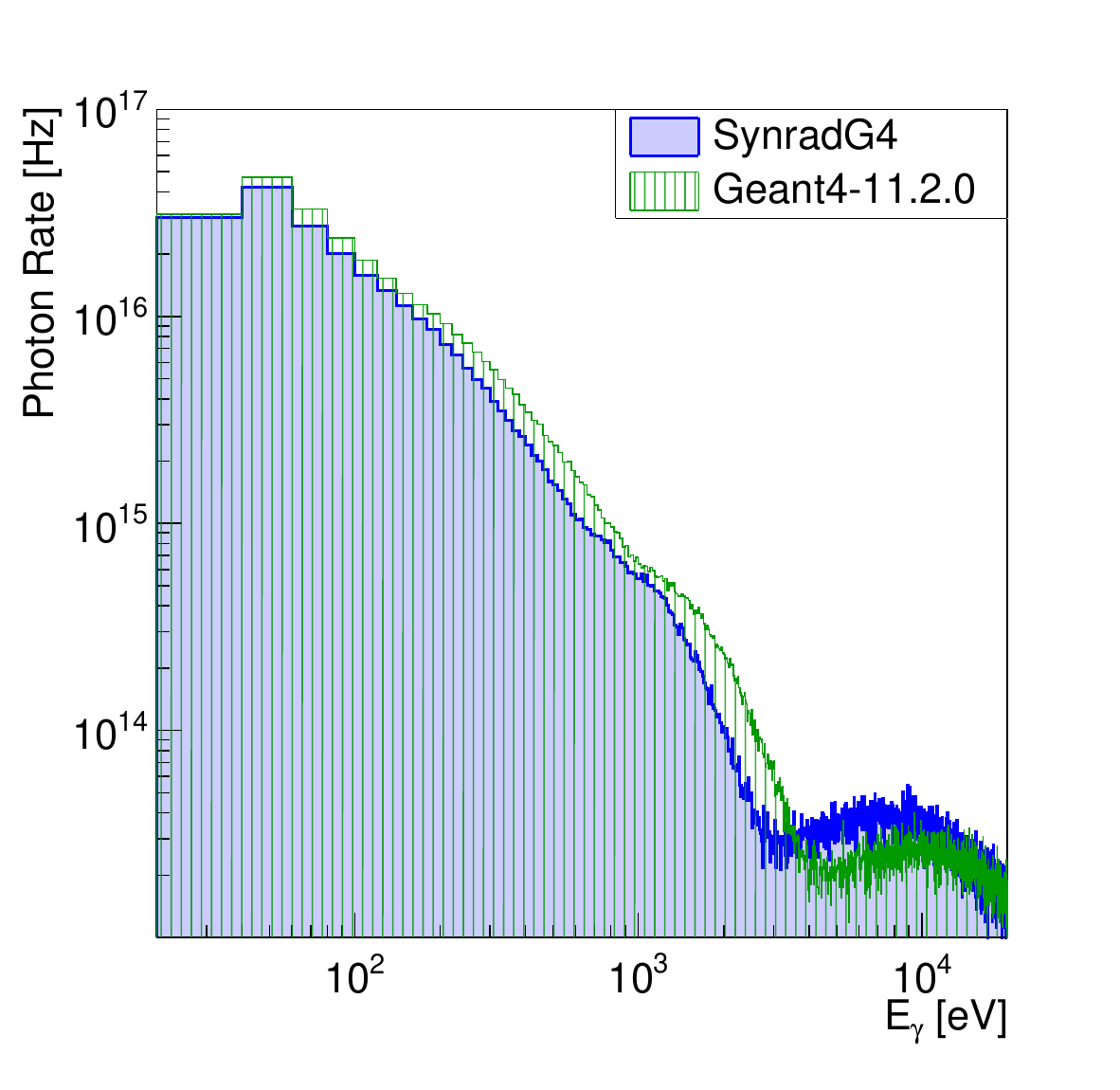}}
\hfill
\subfloat[\label{fig:setup3}Setup-3.]{\includegraphics[width=0.33\linewidth]{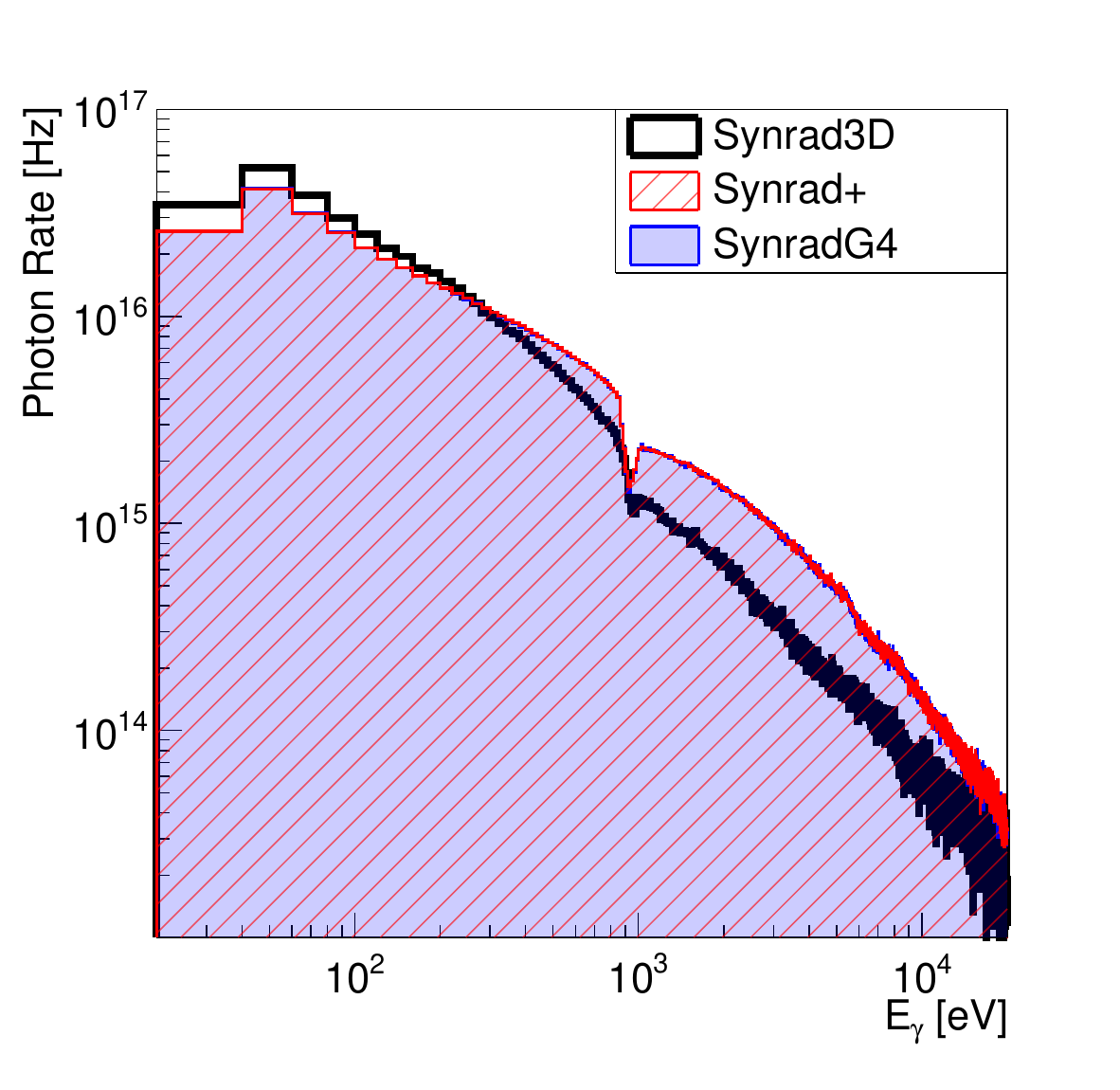}}
\caption{\label{fig:setupResults}Energy spectra of SR photons absorbed between $Z_\mathrm{glob.} = \SI{40}{m}$ and $Z_\mathrm{glob.} = \SI{45}{m}$ for the setups listed in Table~\ref{tab:BenchmarkSetup}. The bin width is \SI{20}{eV}.}

\end{figure*}

Figure~\ref{fig:setupResults} shows the energy spectra of SR photons absorbed between $Z_\mathrm{glob.} = \SI{40}{m}$ and $Z_\mathrm{glob.} = \SI{45}{m}$ for a \SI{1}{A}, \SI{18}{GeV} pencil beam without angular divergence. The ratios quoted below are integrated over the indicated energy ranges.

For smooth surfaces (Setup-1, Fig.~\ref{fig:setup1}), SynradG4 reproduces Synrad+ to within 1\% over the full energy range, as expected, since both codes use the same reflectivity tables and the same treatment of specular reflection. Synrad3D and Geant4-11.2.0 agree with them to within a few percent up to \SI{5}{keV}. At higher energies, where the absorbed-photon rate is lower, the differences follow the reflectivity data used by the codes (Section~\ref{sec:SectionA}): Synrad3D and Geant4-11.2.0, which use the same LBNL data, both predict fewer photons than Synrad+ and SynradG4 between 5 and \SI{13}{keV} (by 20--30\% on average), with common features around \SI{6}{keV} and just below the Cu K edge at \SI{8.98}{keV}, while above \SI{13}{keV} the Geant4-11.2.0 rate is lower than that of the other codes. The total absorbed-photon rates of all four codes agree to within 3\%.

For rough surfaces with specular reflection only (Setup-2, Fig.~\ref{fig:setup2}), the SynradG4 and Geant4-11.2.0 spectra have similar shapes, but Geant4-11.2.0 predicts a higher rate below about \SI{3.5}{keV}: by 10--20\% below \SI{300}{eV}, by 30--45\% between \SI{0.3}{keV} and \SI{1.5}{keV}, rising to a factor of about 2.5 between 2 and \SI{3}{keV}. Between 4 and \SI{10}{keV}, the Geant4-11.2.0 rate is about 35\% lower. In the region where the roughness attenuation causes the spectrum to fall steeply, the Geant4-11.2.0 spectrum is shifted to energies about 25--30\% higher than the SynradG4 spectrum. These differences are consistent with the different exponents of Eqs.~(\ref{eq:eq0}) and (\ref{eq:NC}) discussed in Section~\ref{sec:G4Roughness}. Since the attenuation depends on the product $E_\mathrm{\gamma}\,\sigma\cos\theta_\mathrm{in}$, the weaker attenuation of Eq.~(\ref{eq:NC}) is equivalent to that of Eq.~(\ref{eq:eq0}) at a photon energy lower by a factor of $\sqrt{2}$; the observed shift is somewhat smaller because the smooth-surface reflectivity and the SR spectrum also decrease with energy. With the weaker attenuation, more photons survive the reflections upstream of the scoring region, which increases the rate at intermediate energies. At higher energies, in contrast, a larger fraction of the photons that strike the scoring region at very small grazing angles is reflected out of it instead of being absorbed, which decreases the rate. Differences in the smooth-surface reflectivity (Setup-1) become relevant only above about \SI{5}{keV}.

With diffuse reflection (Setup-3, Fig.~\ref{fig:setup3}), SynradG4 and Synrad+ agree to within about 1\% in all energy ranges, and within statistical fluctuations bin by bin, confirming that the two codes implement the same specular reflection probability, Eq.~(\ref{eq:eq0}), and the same diffuse angular model. Synrad3D predicts up to 25\% more absorbed photons below \SI{300}{eV} and 2 to 3 times fewer above \SI{1}{keV}.

Since SynradG4 and Synrad+ share the same parametrized diffuse model, these differences most likely arise from the different methods used to compute the reflection angles. Synrad3D evaluates the full expression for the diffusely scattered power~\cite{etde_22538279,PhysRevAccelBeams.20.020708}, whereas the parametrization of Eqs.~(\ref{eq:eq1}) and (\ref{eq:eq2}) approximates its high-energy limit $g \gg 1$ (Section~\ref{sec:ReflectionModel}), which is not reached at the lowest photon energies for grazing angles of a few milliradians. Moreover, the approximation of the out-of-plane angle is least accurate for nearly grazing incidence ($1.55 < \theta_\mathrm{in} < \pi/2$, see Fig.~2.32 in Ref.~\cite{Ady:2157666}), which is precisely the regime of SR photons striking the beam-pipe walls. The parametrization could be improved, for example, by using higher-order functions for the fits~\cite{PrivateCommunication2024}, or the angles could be sampled directly from the full Kirchhoff-theory distribution~\cite{PhysRevAccelBeams.20.020708}. 

\begin{table}
    \centering
    \begin{tabular}{l|c}
    \hline\hline
    \multirow{2}{*}{Framework} & Tracking time\\
    & [$\mathrm{\upmu s/photon}$] \\
    \hline
    Synrad3D     &  $\sim 120$\\
    SynradG4     &  $\sim 55$\\
    Synrad+     &  $\sim 8$\\
    \hline\hline
    \end{tabular}
    \caption{\label{tab:SimTime}Tracking time per photon ($E_\mathrm{\gamma} > \SI{30}{eV}$) in Setup-3 for single-core simulations on a MacBook~Pro with an M2~Max processor (\SI{3.68}{GHz}).}
\end{table}

The approximations of the parametrized model come with a substantial gain in speed. As shown in Table~\ref{tab:SimTime}, the tracking time per photon in Setup-3 is about an order of magnitude shorter in Synrad+ than in Synrad3D, which evaluates the infinite sum of the full expression~\cite{etde_22538279}, and SynradG4 is about twice as fast as Synrad3D. The approximated model thus offers a good balance between computational speed and accuracy~\cite{Ady:2157666}, which is essential for the high-statistics SR background studies at the EIC discussed in Section~\ref{sec:SectionA}.

\section{\label{sec:SectionD}SR Background in ePIC}

Using SynradG4, we performed the first end-to-end study of the SR background in the ePIC detector, estimating the background rates induced by SR from the ESR. Here, ``end-to-end'' refers to photon generation in the ESR magnets, fast transport through the $\sim$\SI{50}{m} IR vacuum system (Stage~1), and propagation through the beam-pipe and detector materials (Stage~2). SynradG4 was developed at BNL to enable high-fidelity SR background studies; its baseline source code, with the simplified benchmark geometry, is publicly available~\cite{SynradG4code}.

\begin{figure*}[htbp]
\centering
\includegraphics[width=\linewidth]{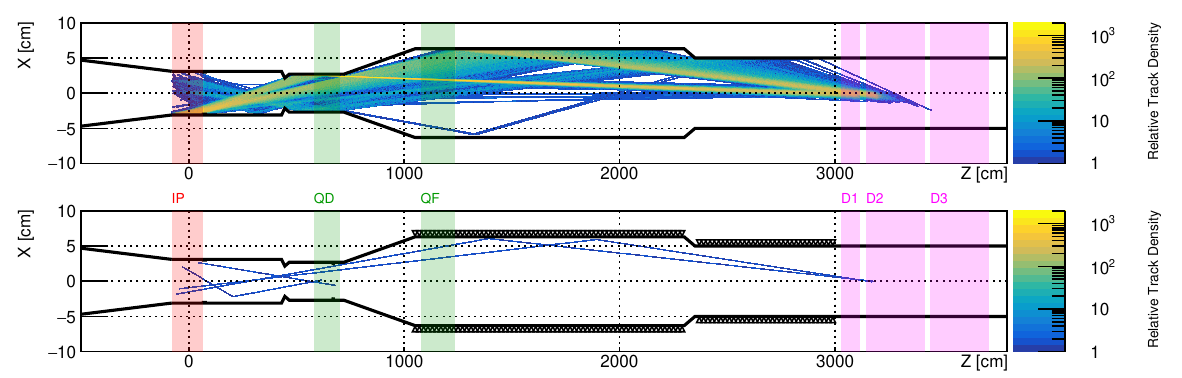}
\caption{\label{fig:SRtracks}Trajectories of SR photons with $E_\mathrm{\gamma} > \SI{10}{keV}$ absorbed on the IP beam pipe. Top: without SR masks; bottom: with SR masks (black open triangles). Red, green, and magenta vertical bands indicate the IP beam pipe, the quadrupole magnets (QD: defocusing, QF: focusing in the horizontal plane), and the dipole magnets (D1--D3), respectively. Solid black lines show the beam-pipe aperture. The electron beam travels from positive to negative $Z$.}

\end{figure*}

The IR beam pipe described in Section~\ref{sec:Introduction} was imported as a tessellated solid, composed of a set of facets, using the \textit{CADMesh} library~\cite{poole2012acad,poole2012fast}. The magnetic field was modeled in the straight section starting from the three nearest dipole magnets (D1--D3; magenta bands in Fig.~\ref{fig:SRtracks}) and includes the two final-focus quadrupole magnets (QD and QF; green bands in Fig.~\ref{fig:SRtracks}). Figure~\ref{fig:SRtracks} shows the trajectories of SR photons with $E_\mathrm{\gamma} > \SI{10}{keV}$ that are absorbed in the IP beam-pipe region (red band). We simulated $10^{7}$~electrons at \SI{18}{GeV}; the SR photons emitted in the magnetic fields were tracked through the beam-pipe vacuum, undergoing multiple reflections described by the diffuse model with the surface parameters of Setup-3.

As shown in Fig.~\ref{fig:SRtracks} (top), most SR photons reach the IP beam pipe after several reflections on the beam-pipe walls. To reduce this rate, we implemented an SR mask inside the last quadrupole magnet (QD) and added mask structures to the beam-pipe surface between \SI{10}{m} and \SI{30}{m} upstream of the IP to intercept the SR photons generated mainly in the nearest dipole magnets (D1 and D2). As shown in Fig.~\ref{fig:SRtracks} (bottom), these SR masks reduce the rate of hard X-rays ($E_\mathrm{\gamma} > \SI{10}{keV}$) hitting the IP beam pipe by more than three orders of magnitude.

We then extended the region of interest and recorded the absorbed SR photons with $E_\mathrm{\gamma} > \SI{30}{eV}$ within $\pm \SI{5}{m}$ of the IP. These photons were propagated through the DD4hep model of the ePIC detector in the eic-shell environment. For the ESR beam parameters listed in Table~\ref{tab:MachineParameters} and $10^{11}$ simulated electrons at \SI{18}{GeV}, the estimated SR background rate in the vertex detector is approximately \SI{1}{THz} without and \SI{1}{GHz} with the SR masks, which demonstrates their effectiveness in mitigating the SR background in the detector. This represents the first official estimate of the SR background for the ePIC detector.

Although the implemented SR masks are relatively \textit{simple} and require further refinement, these results confirm that their development is necessary: according to preliminary estimates, the SR background rate in ePIC should remain below \SI{100}{MHz} to avoid a significant degradation of the tracking performance of the inner ePIC subsystems.

\subsection{Computational Resource Requirements}

Simulating a single event, i.e., tracking one \SI{18}{GeV} electron through the beamline, including SR photon production and transport, currently takes about \SI{10}{ms} on the Jefferson Lab (JLab) Scientific Computing Farm~\cite{JlabIfarm}. A full simulation of the SR background rates at the MHz level in ePIC ($10^{12}$~electrons at \SI{18}{GeV}, corresponding to about \SI{1}{\micro s} of data integration time in the experiment) therefore requires roughly \SI{3e6}{CPU.hours}. Given these high computational demands, we are optimizing the code to accelerate the SR photon transport through the $\sim$\SI{50}{m} long IR vacuum beam pipe.

\section{\label{sec:Conclusion}Conclusions}

We have developed SynradG4, a Geant4-based extension for EIC-specific synchrotron radiation background studies. SynradG4 performs fast SR photon tracking in vacuum with the boundary reflection models of Synrad+ and imports detailed IR vacuum geometries directly from STL meshes. Disabling bulk electromagnetic processes in Stage~1 significantly improves the computational speed while preserving an accurate description of surface scattering. The absorbed photons are then propagated through the full DD4hep detector model, where all relevant X-ray interactions are included.

SynradG4 uses the same specular reflection probability, the Debye--Waller factor of the specularly reflected intensity, and the same parametrized diffuse-reflection model, sampled in the local surface frame, as Synrad+. Benchmark comparisons show that SynradG4 reproduces Synrad+ to within about 1\% for both smooth and rough surfaces, and that it agrees with Synrad3D and Geant4-11.2.0 for smooth surfaces to within a few percent below \SI{5}{keV}. For rough surfaces with specular reflection only, the differences from Geant4-11.2.0 are explained by the convention used for the roughness attenuation factor, which corresponds to a rescaling of the RMS roughness by $\sqrt{2}$. For diffuse reflection, the differences from Synrad3D reflect the approximations of the parametrized angular model, which will be addressed by more detailed models in future work.

Applying SynradG4 to the full EIC IR, we obtained the first SR background estimates for the ePIC detector and demonstrated the effectiveness of upstream SR masks. The tool is publicly available and will be used for further IR optimization and shielding development.

\section{Acknowledgements}

We are grateful to the EIC accelerator, optics, and vacuum groups for their dedicated work in designing this unique machine; to the JLab Scientific Computing Team for providing access to the computing farm, which enabled the high-statistics, CPU-intensive simulations presented here; and to the ePIC colleagues for their contributions to the detector design and the development of the DD4hep model. Special thanks go to D.~Marx (BNL, EIC optics group) for sharing the ESR lattice files, C.~Hetzel (BNL, EIC vacuum group) for providing the 3D models and drawings of the ESR vacuum beam pipe, and W.~Deconinck (University of Manitoba, ePIC software group) for his support in managing the eic-shell code. We also thank M.~Ady (CERN, Synrad+ co-developer) for insightful discussions on Synrad+ and SynradG4, which led to improvements in both frameworks through the resolution of software bugs and an accurate benchmark of the codes, and for providing a stand-alone implementation of the Synrad+ reflection model. Additionally, we acknowledge D.C.~Sagan and G.H.~Hoffstaetter de Torquat (Cornell University) for introducing the Bmad framework used in the Synrad3D simulation, and M.K.~Sullivan (SLAC), H.~Witte, and E.-C.~Aschenauer (BNL) for their valuable ideas, assistance, and constructive discussions.

This work was supported by Brookhaven Science Associates, LLC under Contract No.~DE-SC0012704 with the U.S. Department of Energy.

\bibliography{mybibfile}

\end{document}